\documentclass[a4paper]{aipproc}

\layoutstyle{6x9}
  
\begin{document}

\title 
[Recent measurements with the out-of-plane spectrometer system at MIT-Bates]
{Recent measurements with the out-of-plane spectrometer system at MIT-Bates}


\author{Simon \v Sirca, for the OOPS Collaboration}{
  address={MIT, Laboratory for Nuclear Science, 77 Massachusetts Avenue,
    Cambridge, MA 02139, USA},
  email={sirca@mitlns.mit.edu},
  thanks={for the OOPS Collaboration}	
}

\copyrightyear{2001}

\begin{abstract}
The recent experimental program  with the out-of-plane spectrometer
system (OOPS) at MIT-Bates encompassed an extensive set of
$\mathrm{d}(\vec\mathrm{e},\mathrm{e}'\mathrm{p})$ measurements,
investigations of the $\mathrm{N}\to\Delta$ transition
using $\mathrm{p}(\vec\mathrm{e},\mathrm{e}'\mathrm{p})\pi^0$ and
$\mathrm{p}(\vec\mathrm{e},\mathrm{e}'\pi^+)\mathrm{n}$ reaction
channels, and studies of virtual Compton scattering (VCS)
$\mathrm{p}(\mathrm{e},\mathrm{e}'\mathrm{p})\gamma$
below the pion threshold.  Preliminary results are presented.
\end{abstract}

\date{\today}

\maketitle

\subsection{Measurements of $\mathbf{d}(\vec\mathbf{e},\mathbf{e}'\mathbf{p})$
in the dip region}

Early measurements of unpolarised responses for deuteron
electro-disintegration have yielded a substantial, but 
inconsistent body of data which could not be adequately 
described by theoretical models (see \cite{newobs} for a review).
To provide a richer input to theories, the deuteron program
at Bates has been dedicated to separations of interference
responses in a variety of kinematical settings.  This could be
achieved by simultaneous out-of-plane detection of ejected hadrons
about the momentum transfer, thereby minimising systematic uncertainties
and allowing for separation of both asymmetries and absolute responses
\cite{oopsnim1,oopsnim2}.  With this novel technique, competing effect
in deuteron electro-disintegration (final-state interactions (FSI),
meson-exchange currents (MEC), isobar configurations (IC),
and relativistic corrections (RC)) can be probed precisely and selectively.

\begin{figure}[hbt]
  \includegraphics[height=.25\textheight]{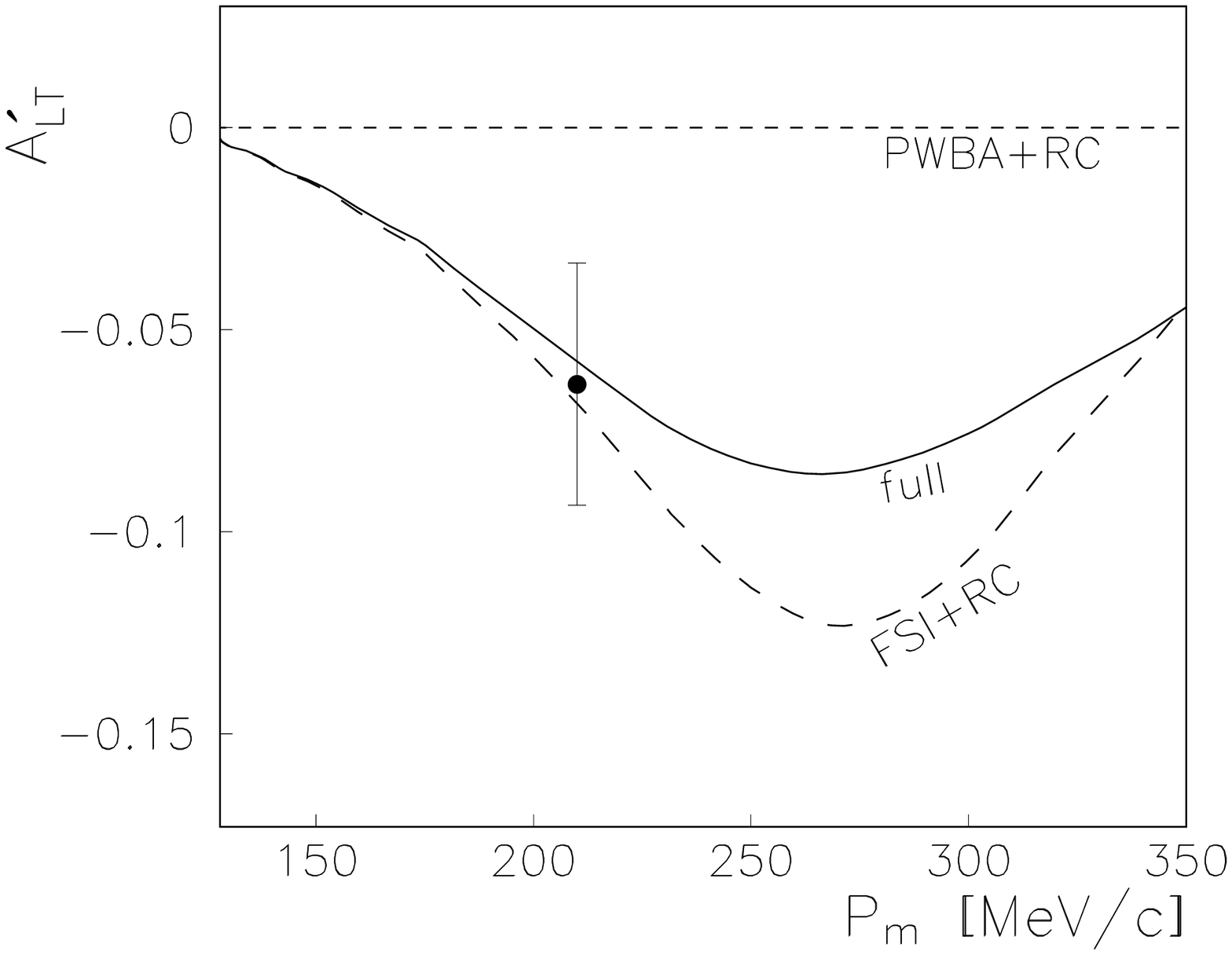}
  \includegraphics[height=.24\textheight]{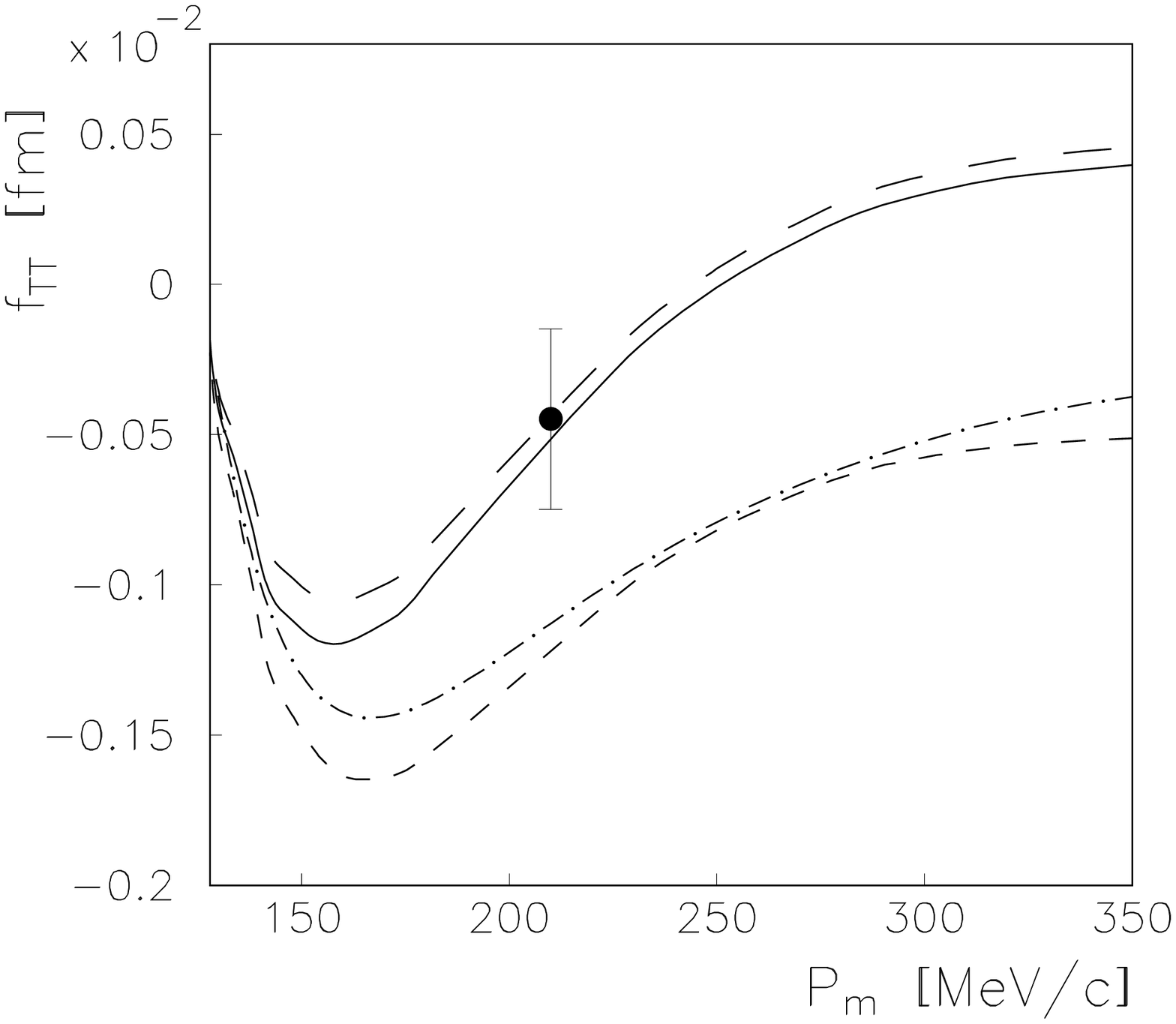}
  \caption{Preliminary results for 
  the asymmetry $A_\mathrm{LT}'$ and the $f_\mathrm{TT}$
  response as functions of $p_\mathrm{m}$.  Left panel: calculations
  of ref.~\protect\cite{aren} with ingredients marked in the figure.
  Right panel: calculations of ref.~\protect\cite{aren}: relativistic
  PWBA+FSI+RC (dashed), PWBA+FSI+MEC+IC (long-dashed), and full (solid),
  and of ref.~\protect\cite{tjon}: PWBA+FSI (dashed-dotted).
  Note that both of these observables require out-of-plane detection.}
  \label{altp_ftt_dip}
\end{figure}

Early deuteron electro-disintegration studies at MIT-Bates were
focused on the quasi-elastic region \cite{dolfini,jordan}.
We report here on the measurements of
$\mathrm{d}(\vec\mathrm{e},\mathrm{e}'\mathrm{p})$ 
in the dip region, where the asymmetries $A_\mathrm{LT}$,
$A_\mathrm{LT}'$, and $A_\mathrm{TT}$, as well as the responses
$f_\mathrm{LT}$, $f_\mathrm{LT}$, and $f_\mathrm{TT}$ were measured
at a beam energy of $800\,\mathrm{MeV}$, a four-momentum transfer
of $Q^2=0.15\,\mathrm{GeV}^2$, and a missing momentum of
$210\,\mathrm{MeV/c}$.  Preliminary results on the asymmetry
$A_\mathrm{LT}'$ and the $f_\mathrm{TT}$ response are given
in figure~\ref{altp_ftt_dip}.
The asymmetry $A_\mathrm{LT}'$ reflects the imaginary part of the
$\mathrm{LT}$ interference term, and is thus highly sensitive to FSI.
The $f_\mathrm{TT}$ response is very sensitive to an accurate
inclusion of MEC and IC, and exhibits almost no dependence on relativity,
contrary to $A_\mathrm{LT}$ and $f_\mathrm{LT}$.  These results
have been submitted for publication in Phys. Rev. Lett.

These measurements were performed with pulsed ($\simeq 1\,\%$-duty-factor)
beam, while the recent availability of the high-duty-factor extracted beam
has challenged the OOPS Collaboration to continue the pursuit of
the out-of-plane program, to extend it to higher missing momenta
and into the $\Delta$-region.  At higher energy transfers
and higher missing momenta, an enhanced sensitivity to relativistic
effects in $A_\mathrm{LT}$ and $f_\mathrm{LT}$, and to MEC and IC
in $A_\mathrm{TT}$ and $f_\mathrm{TT}$ is expected.  In the same
kinematical region, the $A_\mathrm{LT}'$ and $f_\mathrm{LT}'$ will
improve our understanding of $\Delta-\mathrm{N}$ interactions
in the final state.  In addition, these studies can be
complemented with an enlarged set of polarisation observables,
and can be performed with a significant decrease in experimental
uncertainties.

\subsection{Upgrade and commissioning of the MIT-Bates South Hall and OOPS}

For the two most recent experimental efforts at MIT-Bates which 
required a high-duty-factor beam, major instrumental developments
have taken place in the South Hall of the facility.
A gantry support system allowing for out-of-plane positioning of two
OOPS modules has been completed, and the fourth OOPS module has been
commissioned.  The OHIPS spectrometer has been upgraded
for a momentum bite of $14\,\%$ and outfitted with a new vertical
drift-chamber and additional scintillators and lead-glass detectors.
Substantial modifications were made to the beam-line and readout
electronics to conform to the CW beam extracted from
the South-Hall storage ring \cite{batesreport}.

\subsection{Studies of the $\mathrm{N}\to\Delta$ transition}

Studies of the $\gamma^\star\mathrm{N}\to\Delta$ transition
using OOPS have a long tradition \cite{nd_props}.  They were originally
aimed at precise extractions of the $\mathrm{E2}/\mathrm{M1}$ and
$\mathrm{C2}/\mathrm{M1}$ multipole amplitude ratios which 
quantify the deviation of the nucleon or the $\Delta$ from a spherical
shape such as that assumed in the naive non-relativistic quark model
or in models with a spherically-symmetric pion field.  The initial
effort to disentangle the quadrupole amplitudes from the dominating
magnetic dipole component has now been augmented by more
complete investigations.  The $\mathrm{LT}$, $\mathrm{LT}'$, and
$\mathrm{TT}$ interference responses and the corresponding asymmetries
have been isolated in wide kinematical regions.
The $\mathrm{p}(\mathrm{e},\mathrm{e}'\mathrm{p})\pi^0$ process
has been studied in a range of invariant masses in the vicinity
of the $\Delta$-resonance peak to explore the $W$-dependence of
resonance and background contributions.  A handle on the isospin
structure of the $\mathrm{N}\to\Delta$ transition has been obtained
by measurements of the concurrent
$\mathrm{p}(\mathrm{e},\mathrm{e}'\mathrm{n})\pi^+$ process.
Angular distributions in $\theta_\mathrm{pq}^\star$
(or $\theta_{\pi\mathrm{q}}^\star$) of protons (or $\pi^+$) have been
measured to allow for a multipole decomposition of the responses.

\begin{figure}[hbt]
  \includegraphics[height=.3\textheight]{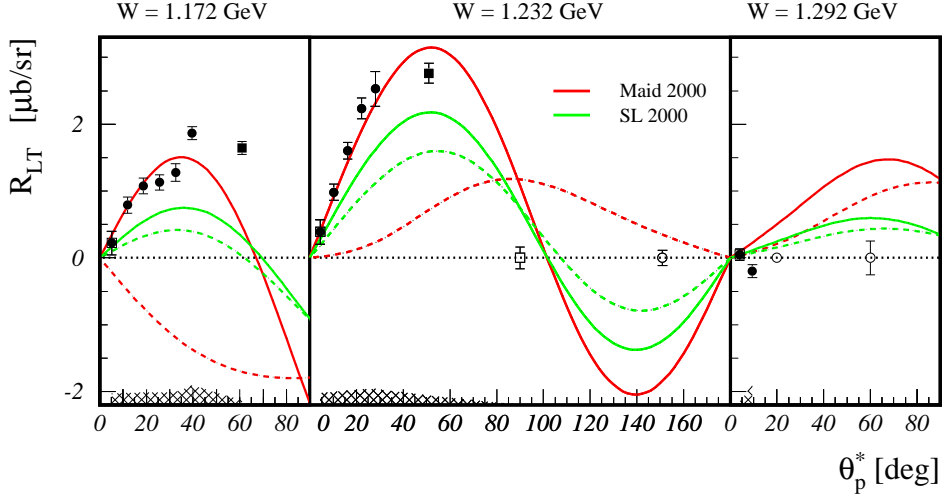}
  \caption{The $R_\mathrm{LT}$ response at $Q^2=0.126\,\mathrm{GeV}^2$
  and $W=1172$, $1232$ and $1292\,\mathrm{MeV}$. The open square
  and the open circles denote the expected
  statistical uncertainties for the 2001 run.  The shaded areas
  represent the systematic uncertainties.  Full circles and squares:
  previously taken data \protect\cite{mertz,kunz}.  Model predictions are
  by Sato and Lee \protect\cite{satolee} and MAID \protect\cite{maid}.
  The dashed curves correspond to calculations without the $\mathrm{C2}$
  contribution.}
  \label{rlt}
\end{figure}

\begin{figure}[hbt]
  \includegraphics[height=.3\textheight]{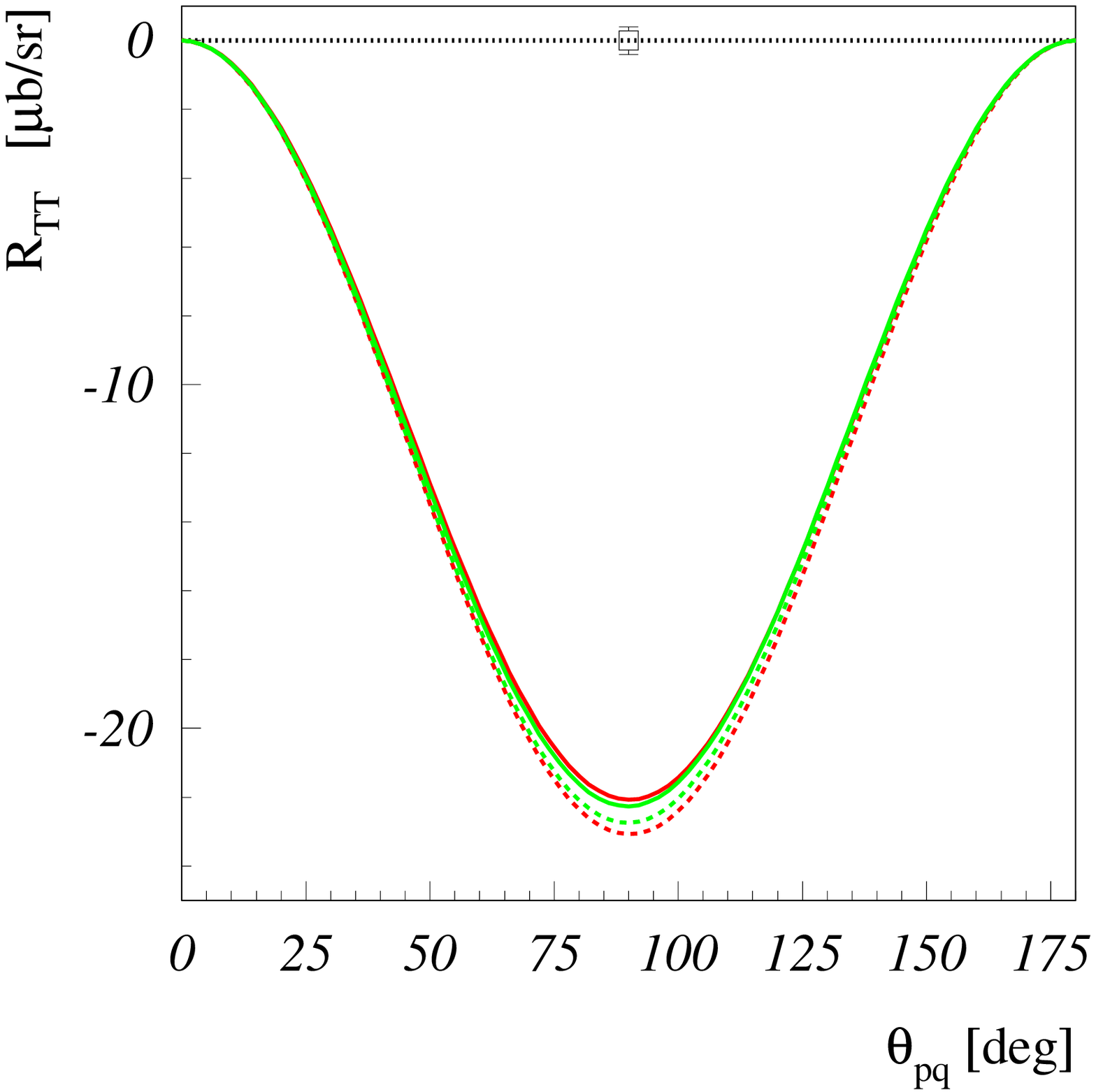}
  \includegraphics[height=.3\textheight]{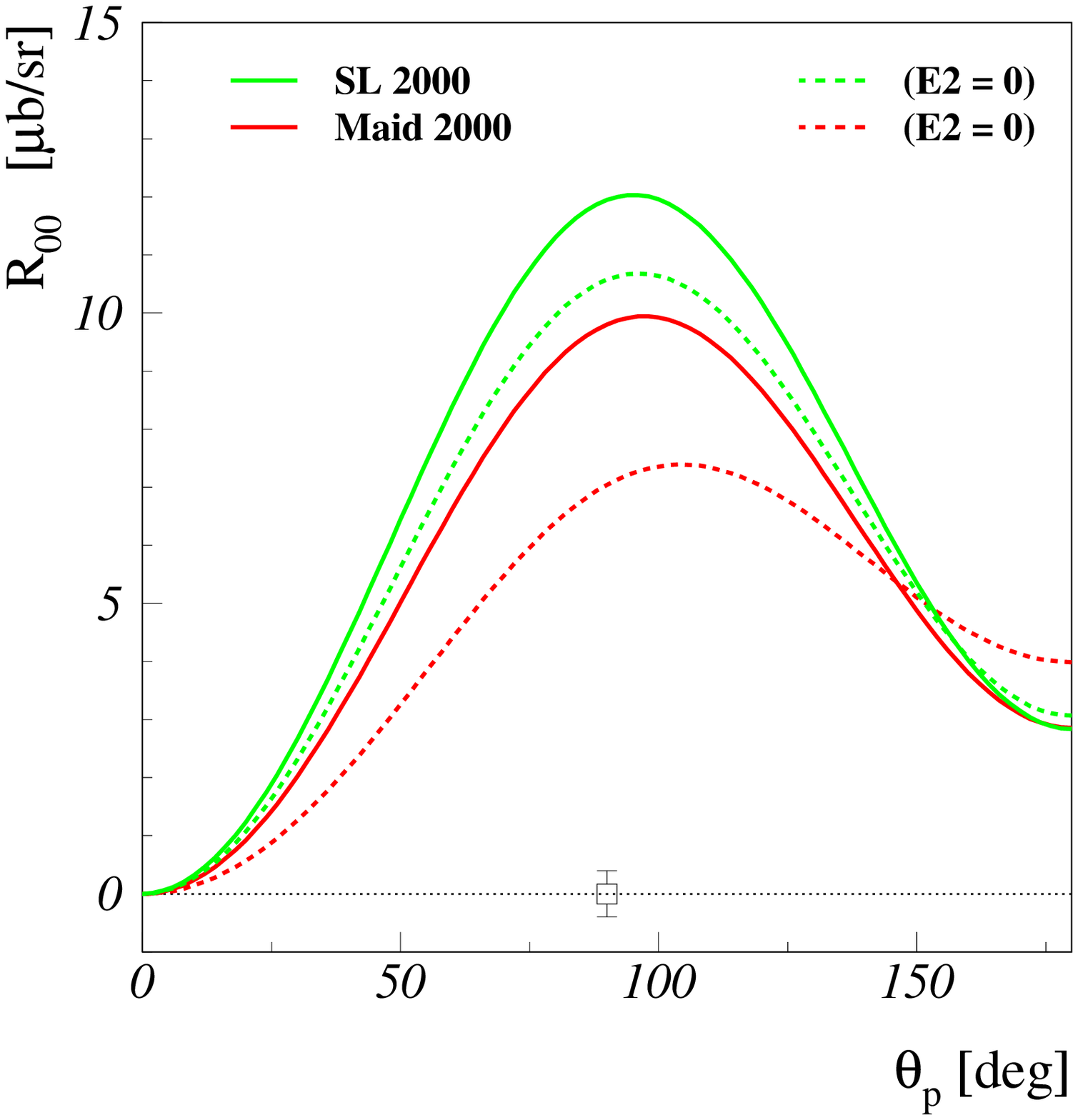}
  \caption{Expected statistical uncertainties for $R_\mathrm{TT}$
  and $R_{00}$ (see text) in the $\pi^0$-channel at $W=1232\,\mathrm{MeV}$
  and $Q^2=0.126\,\mathrm{GeV}^2$.  The empty square in the right panel
  corresponds to about a third of the data taken in the run of 2001.
  (Data at $\theta_\mathrm{pq}^\star=151^\circ$ was also taken.)
  For curves, see caption to figure~\protect\ref{rlt}.}
  \label{rtt_r00}
\end{figure}

During the 2001 run, the OOPS system was operated in the full,
four-module setup for the first time, with a beam energy of
$950\,\mathrm{MeV}$ at high duty-factor currents of up to
$10\,\mu\mathrm{A}$.  Two modules mounted on the gantry,
in conjunction with the two in-plane modules, allowed
for a decomposition of all unpolarised response functions,
including $R_\mathrm{TT}$.  Running at forward angles and low proton
momenta permitted a significant extension of the range of angular
distributions.  Figure~\ref{rlt} shows the $R_\mathrm{LT}$ response
which offers a most precise test of the existing phenomenological
models.  It is highly sensitive to the poorly-known resonance
interference term $\mathrm{Re}(S_{1+}^\star M_{1+})$, except at
the particular angle of $\theta_\mathrm{pq}=90^\circ$ where it
is suppressed in favour of the maximal sensitivity to the background
term $\mathrm{Re}(S_{0+}^\star M_{1+})$.
The $R_\mathrm{TT}$ response has never been measured before
(see left panel of figure~\ref{rtt_r00}).  This pure out-of-plane
response is important in calibrating the absolute strength of the
dominant $M_{1+}$ transition multipole.  The particular combination
of responses at $\theta_\mathrm{pq}^\star=0^\circ$ (in parallel
kinematics) and at an emission angle of $90^\circ$,
$R_{00} \equiv (R_\mathrm{T}+\varepsilon_\mathrm{L}R_\mathrm{L})(90^\circ)
              + R_\mathrm{TT}(90^\circ)
              -(R_\mathrm{T}+\varepsilon_\mathrm{L}R_\mathrm{L})(0^\circ)$
will also be extracted, since it exhibits a unique sensitivity to the
$\mathrm{E2}$ amplitude (see right panel of figure~\ref{rtt_r00}).
Excellent beam conditions also enabled us to carry out a 
measurement of interference responses and cross-sections in the
$\pi^+$-channel at $W=1232\,\mathrm{MeV}$, $Q^2=0.126\,\mathrm{GeV}^2$,
and $\theta_{\pi\mathrm{q}}^\star=44.5^\circ$, which is being
analysed.  The combined results of both reaction channels will yield
information on the isospin structure of the $\mathrm{N}\to\Delta$ transition.

\subsection{Virtual Compton scattering}

The Virtual Compton scattering experiment \cite{vcsprop} was one
of the highlights of this year's extracted-beam program.  The process
$\mathrm{p}(\mathrm{e},\mathrm{e}'\mathrm{p})\gamma$ has been
measured at five beam energies, corresponding to the outgoing-photon
energies ranging from $28$ to $115\,\mathrm{MeV/c}$, at four-momentum
transfer of $Q^2=0.05\,\mathrm{GeV}^2$.  The physical goal was
to measure the generalised polarisabilities of the proton at low momentum
transfers and test several competing theories with great accuracy
and minimal systematical uncertainties.  The OOPS setup is excellently
matched to access these observables at low momentum transfers,
in particular due to a great out-of-plane suppression of the Bethe-Heitler
background.  Contrary to in-plane experiments which are primarily
sensitive to the magnetic generalised polarisabilities, out-of-plane
detection uniquely allows for a clear isolation of the electric part.
A detailed analysis of the data is underway.

\end{document}